\documentclass[useAMS,usenatbib,usegraphicx]{mn2e}

\usepackage{url}
\usepackage[varg]{txfonts}
\usepackage{times}
\usepackage{amssymb}
\usepackage{xspace}
\usepackage{graphicx}

\def\dark{Dark Cosmology Centre, Niels Bohr Institute, University of
Copenhagen, Juliane Maries Vej 30, 2100 Copenhagen {\O}, Denmark}

\def\nbia{Niels Bohr International Academy, Niels Bohr Institute,
University of Copenhagen, Blegdamsvej 17, 2100 Copenhagen {\O},
Denmark}

\def\ucsb{Dept. of Physics, University of California, Santa Barbara,
CA 93106, USA}

\def\kapteyn{Kapteyn Astronomical Institute, University of Groningen,
P.~O.~Box 800, 9700 AV Groningen, The Netherlands}

\def\vienna{Institut f\"{u}r Astrophysik, Universit\"{a}t
Wien, T\"{u}rkenschanzstra{\ss}e 17, 1180 Wien, Austria}

\newcommand{\aj}{AJ}       
\newcommand{\apj}{ApJ}     
\newcommand{\apjl}{ApJ}    
\newcommand{\apjs}{ApJS}   
\newcommand{\mnras}{MNRAS} 
\newcommand{\aap}{A\&A}    
\newcommand{\araa}{ARAA}   
\newcommand{\pasp}{PASP}   

\newcommand{\nat} {Nature}

\newcommand {\kms} {\ifmmode  \,\rm km\,s^{-1} \else $\,\rm km\,s^{-1}$ \fi }
\newcommand {\kpc} {\ifmmode  {\rm kpc}  \else ${\rm  kpc}$ \fi  }  
\newcommand {\pc} {\ifmmode  {\rm pc}  \else ${\rm pc}$ \fi  }  
\newcommand {\Msun} {\ifmmode {M_{\odot}} \else ${M_{\odot}}$ \fi} 
\newcommand {\Zsun} {\ifmmode {\rm Z_{\odot}} \else ${\rm Z_{\odot}}$ \fi} 
\newcommand {\yr} {\ifmmode yr^{-1} \else $yr^{-1}$ \fi} 
\newcommand {\hMsun} {\ifmmode h^{-1}\, M_{\odot} \else $h^{-1}\, M_{\odot}$ \fi}

\usepackage[usenames]{color}

\newcommand{\simf}{x} 
\newcommand{\Mcutoff}{M_{\mathrm{low}}} 
\newcommand{\logM}{\log_{10} \Ms/\Msun}
\newcommand{\simfJ}[1]{\simf_{\mathrm{#1}}}

\newcommand{\slope}{\gamma}

\newcommand{\Reff}{R_{\mathrm{e}}}

\newcommand{\fDM}{f_{\mathrm{DM}}}

\newcommand{\MoLs}{\Upsilon_{\star}}
\newcommand{\Ms}{M_{\star}}
 
\newcommand{\vvir}{v_{\mathrm{vir}}} 
\newcommand{\cg}{c_{-2}} 
\newcommand{\qh}{q_{\rm h}}

\title[IMF slope and cut-off in ETGs] {A low-mass cut-off near the
  hydrogen burning limit for Salpeter-like initial mass functions in
  early-type galaxies}

\author[Barnab\`e et al.]{%
  Matteo~Barnab\`e,$^{1,2}$
  Chiara~Spiniello,$^{3}$
  L\'eon~V.~E.~Koopmans,$^{3}$
  Scott~C.~Trager,$^{3}$
\newauthor{%
  Oliver~Czoske,$^{4}$
  and Tommaso~Treu$^{5}$}
\smallskip\\
  $^1$\dark\\
  $^2$\nbia\\
  $^3$\kapteyn\\
  $^4$\vienna\\
  $^5$\ucsb\\
}

\begin{document}

\date{Accepted 2013 August 16.  Received 2013 August 8; in original form 2013 June 11}

\maketitle

\label{firstpage}

\begin{abstract} 
  We conduct a detailed investigation of the properties of the stellar
  initial mass function (IMF) in two massive early-type lens galaxies
  with velocity dispersions of $\sigma \simeq 245 \kms$ and $\sigma
  \simeq 325 \kms$, for which both \emph{HST} imaging and X-Shooter
  spectra are available.  We compare the inferences obtained from two
  fully independent methods: (i) a combined gravitational lensing and
  stellar dynamics (L\&D) analysis of the data sets employing
  self-consistent axisymmetric models, and (ii) a spectroscopic simple
  stellar population (SSP) analysis of optical line-strength indices,
  assuming single power-law IMFs.  The results from the two approaches
  are found to be in agreement within the 1-$\sigma$
  uncertainties. Both galaxies are consistent with having a Salpeter
  IMF (power-law slope of~$\simf = 2.35$), which is strongly favoured
  over a Chabrier IMF ($\simf = 1.8$), with probabilities inferred
  from the joint analysis of~$89\%$ and $99\%$,
  respectively. Bottom-heavy IMFs significantly steeper than Salpeter
  ($\simf \ge 3.0$) are ruled out with decisive evidence (Bayes factor
  $B > 1000$) for both galaxies, as they exceed the total mass derived
  from the L\&D constraints.  Our analysis allows, for the first time,
  the inference of the low-mass cut-off of the IMF ($\Mcutoff$).
  Combining the joint L\&D and SSP analyses of both galaxies, we infer
  an IMF slope of $\simf = 2.22 \pm 0.14$, consistent with Salpeter
  IMF, and a low-mass limit $\Mcutoff = 0.13 \pm 0.03 \, \Msun$, just
  above the hydrogen burning limit.
\end{abstract}




\section{Introduction}
\label{sec:introduction}

The stellar initial mass function is a fundamental quantity to
understand the evolution and structure of galaxies, since it is
crucial in determining many key properties of stellar populations and
galaxies, such as star formation rates and galaxy stellar masses.
While it has long been assumed that the IMF is universal, given that
the IMF appears to be largely invariant throughout the Local Group
\citep{Kroupa2001, Chabrier2003, Bastian2010, Kroupa2013book}, in
recent years numerous extra-galactic studies have provided mounting
evidence that the steepness of the IMF profile might vary with galaxy
mass, stellar velocity dispersion or morphological type (e.g.\
\citealt{Trager2000b, Graves2009, Treu2010, vanDokkum-Conroy2010,
  Auger2010imf}; \citealt[hereafter B12]{Barnabe2012};
\citealt{Geha2013, Dutton2013-SW}). In particular, there are strong
indications that massive elliptical galaxies are characterized by a
classic single power-law \citet{Salpeter1955} IMF ($dN/dm \propto
m^{-\simf}$ with slope $\simf = 2.35$, where $N$ is the number of
stars of mass $m$), possibly becoming steeper with increasing total
galaxy mass (see e.g.\ \citealt{Grillo2009};
\citealt{vanDokkum-Conroy2010}; \citealt{Barnabe2009, Barnabe2011};
\citealt{Cappellari2012}; \citealt{Spiniello2011, Spiniello2012,
  vanDokkum-Conroy2012}; \citealt{Conroy-vanDokkum2012b};
\citealt{Sonnenfeld2012}; \citealt{Ferreras2013};
\citealt{Tortora2013}).

Understanding the trend in the IMF slope is necessary to determine how
the dark matter content of galaxies changes with mass, and thus to
provide important insight into galaxy evolution mechanisms. Several
works \citep[e.g.][]{Padmanabhan2004, Tortora2009, Graves-Faber2010,
  Barnabe2011} have shown that, under the assumption of a universal
IMF, the inferred dark matter fraction within the inner regions of
early-type galaxies (ETGs) increases with the total mass. However, it
remains unclear whether (and how much) this effect can be countered by
the observed variation in the IMF slope, and consequently to what
extent (if at all) the actual dark matter content of ETGs scales with
total mass.

The combination of gravitational lensing with stellar (or gas)
dynamics has been proven to be a very robust and powerful method to
investigate the mass structure of both early- and late-type galaxies
beyond the local Universe \citep[e.g.][]{Treu-Koopmans2004,
  Jiang-Kochanek2007, Czoske2008, Koopmans2009, Barnabe2010,
  Suyu2012}. When lensing and kinematic data of sufficient quality are
available, this technique can also make it possible to put firm
constraints on the stellar mass of the analyzed galaxy (see
\citealt{vandeVen2010, Dutton2011-SW}; B12), which can then be
compared to the results of SSP modelling to draw inferences for the
shape of the IMF of the galaxy. In this Letter we follow an analogous
approach: by taking advantage of high-quality X-Shooter spectroscopic
observations, we perform a fully self-consistent joint L\&D analysis
of two massive lens ETGs: SDSS\,J0936$+$0913 ($\sigma = 243 \pm 11$
$\kms$ at~$z = 0.164$) and SDSS\,J0912$+$0029 ($\sigma = 326 \pm 12$
$\kms$ at~$z = 0.190$), hereafter J0936 and J0912, respectively. The
inferences for the stellar masses are then compared against the
corresponding results from a completely independent method, viz., a
spectroscopic SSP study of the same two systems based on line-strength
indices analysis, thus yielding secure constraints on the IMF slope
and, for the first time, on the IMF low-mass cut-off. The value of
$\Mcutoff$ is an essential but often unappreciated parameter when
determining the stellar mass-to-light ratio $\MoLs$ from stellar
population evolutionary codes. In fact, different codes adopt a
different choice for $\Mcutoff$, implying different results for
$\MoLs$ (e.g., \citealt{Conroy-vanDokkum2012a}, hereafter CvD12, use
$0.08 \Msun$, \citealt{Bruzual-Charlot2003} and \citealt{Vazdekis2012}
adopt $0.10 \Msun$, while all models based on the {\sl Padova 2000}
isochrones use $0.15 \Msun$).


\section{Data set}
\label{sec:observations}

Both galaxies were originally observed with \emph{HST} as part of the
Sloan Lens ACS Survey (SLACS). We refer to \citet{Bolton2008a} and
\citet{Auger2009} for a detailed presentation of all the SLACS
spectrophotometric data sets. For the analysis presented here, we make
use of the high-resolution \emph{F814W} images of the lens systems. An
elliptical B-spline model of the surface brightness distribution of
the deflector is used to produce the galaxy-subtracted image that is
used for the lensing modelling \citep[see][]{Bolton2008a}.

For both systems, UVB--VIS X-Shooter spectra with signal-to-noise high
enough to perform SSP analysis ($\mathrm{S/N} > 50$) were obtained as
part of the X-Shooter Lens Survey (XLENS, \citealt{Spiniello2011}).
X-Shooter observations of J0912 and J0936 were carried out during two
runs between 2011 and 2013 in slit mode (UVB: $\mathrm{R} = 3300$ with
$1\farcs6\times 11\farcs$ slit; VIS: $\mathrm{R} = 5400$, with
$1\farcs5 \times 11\farcs$ slit).  The total exposure time on target
was $\sim 2500$ sec for J0912 and $\sim 5000$ sec for J0936, with a
typical seeing for both observations of $\sim0\farcs65$.  The data
were reduced using the ESO X-Shooter pipeline v1.5.7.  Stellar
kinematic parameters were measured with the Penalized Pixel Fitting
code of \citet{Cappellari-Emsellem2004}.
 

\section{Simple stellar population modelling}
\label{sec:SSP}

For our analysis we use the SSP models of CvD12. To constrain the IMF
slope, age and [$\alpha$/Fe] of the stellar populations, we follow the
approach described by \citet{Spiniello2013}.  We use the nearly
IMF-independent indicators H$\beta$, Mg$b$, Fe5270, and Fe5335 to
constrain age and [$\alpha$/Fe], while the combination of CaH1 and
four TiO features are used to measure the low-mass end of the IMF. We
compare the indices of each galaxy with grids of SSPs spanning a range
of ${\log_{10} (\mathrm{age/Gyr})} = [0.8, 1.15]$ and [$\alpha$/Fe] $
= [-0.2, +0.4]$ dex for different values of the IMF slope ($x = 1.8 -
3.5$), which is assumed to be a single power-law.  In CvD12 the
abundance variations of single elements are implemented at fixed
[Fe/H], which implies that the total metallicity~$Z$ varies from model
to model.
For each model we calculate the $\chi^{2}$ and obtain a probability
distribution function (PDF) via the likelihood function and assuming
uniform priors over the above described ranges.  We then marginalize
over age and [$\alpha$/Fe] to obtain the best-fitting slope of the IMF
and its uncertainty for each system: $\simfJ{J0936} = 2.10 \pm 0.15$
and $\simfJ{J0912} = 2.60 \pm 0.30$.

\begin{table*}
  \centering
  \begin{minipage}{0.76\linewidth}
    \caption{SSP analysis results and inferred stellar masses for the
      two different $\Mcutoff$ values.}
    \label{tab:ssp}
    \begin{tabular}{lccccc} 
\hline
system  & $\log_{10} (\mathrm{age/Gyr})$ & [$\alpha$/Fe] & IMF slope & $M_{\star}$ with $\Mcutoff = 0.115 \Msun$ & $M_{\star}$ with $\Mcutoff = 0.08 \Msun$ \\
\hline
J0936 	& $0.9 \pm 0.05$  & $0.05 \pm 0.02$ & $2.1\pm 0.15$ & $(3.01 \pm 0.03) \times 10^{11} \Msun$ & $(3.56 \pm 0.03) \times 10^{11} \Msun$  \\
J0912 	& $1.1 \pm 0.05$  & $0.10 \pm 0.03$ & $2.6\pm 0.30$ & $(1.28 \pm 0.04) \times 10^{12} \Msun$ & $(1.78 \pm 0.04) \times 10^{12} \Msun$  \\
\hline
     \end{tabular}
  \end{minipage}
\end{table*}

To determine~$\MoLs$, we use the isochrones at solar [Fe/H] from the
state-of-the-art stellar evolution code DSEP (Dartmouth Stellar
Evolution Program, \citealt{Chaboyer2001}), selecting IMF slope, age,
and [$\alpha$/Fe] inferred from the line-strength analysis (see
Table~\ref{tab:ssp}).  Each $\MoLs$ includes the contribution from
stellar remnants and gas ejected from stars at the end of their
life-cycles.
The lowest mass limit in the DSEP isochrones is $\Mcutoff = 0.115 \,
\Msun$.  We extrapolate the luminosities for lower mass down to the
hydrogen burning limit of $\Mcutoff= 0.08 \, \Msun$
\citep[e.g.][]{Kumar1963, Grossman1970}, using a spline.  We note that
stars below $\sim 0.115 \, \Msun$ remain mostly invisible in current
spectral lines for any reasonable IMF slope (CvD12) and therefore we
can move~$\Mcutoff$ without any real impact on the spectra and on the
line-strength measurements.  Finally, we use the luminosities derived
in \citet{Auger2009} to translate these numbers into stellar
masses. The results are presented in Table~\ref{tab:ssp}.


\section{Gravitational lensing and stellar kinematics modelling}
\label{sec:mass-model}

We carry out an in-depth analysis of the mass and dynamical structure
of the two lens systems by employing the fully Bayesian
\textsc{cauldron} code (detailed by \citealt{Barnabe-Koopmans2007} and
B12), which is designed to conduct a self-consistent combined
modelling of both the lensing and kinematic constraints.

We adopt a flexible two-component axially-symmetric mass model for the
lens galaxy. For the dark matter component, motivated by the findings
of cosmological simulations, we use a generalized Navarro-Frenk-White
(gNFW) halo characterized by four free parameters, namely the inner
density slope~$\slope$, the axial ratio~$\qh$, the halo concentration
parameter~$\cg$ and the virial velocity~$\vvir$ (i.e. the circular
velocity at the virial radius). The density profile of the luminous
mass component is obtained by deprojecting the multi-Gaussian
expansion (MGE) fit to the observed surface brightness distribution of
the galaxy, a technique that has become the \emph{de facto} standard
in state-of-the-art dynamical studies of galaxies \citep[e.g., the
  ATLAS$^{\mathrm{3D}}$ project,][]{Cappellari2011}. The free
parameter~$\MoLs$, or equivalently the total stellar mass~$\Ms \equiv
\MoLs L_{\mathrm{tot}}$, sets the normalization of the luminous
distribution.

This mass profile is then used to simultaneously model both the
lensing data set, by employing a pixelated source reconstruction
method, and the stellar kinematic observables, i.e.\ the quantity
$v_{\mathrm{rms}}(R)= \sqrt{v_{\mathrm{rot}}^2(R) + \sigma^2(R)}$,
where $v_{\mathrm{rot}}$ and $\sigma$ denote the line-of-sight
projected stellar rotation velocity and velocity dispersion,
respectively. The construction of the dynamical model is based on the
flexible Jeans anisotropic MGE technique \citep[JAM,
  see][]{Cappellari2008}, which allows for orbital anisotropy,
specified by the meridional plane anisotropy parameter~$b =
\sigma^{2}_{R}/\sigma^{2}_{z} = 1/(1 - \beta_{z})$.

The adopted prior distributions for the six free parameters are
presented in Table~\ref{table:par}, together with the one-dimensional
marginalized posterior PDFs derived from the combined L\&D analysis.
As discussed by B12, we choose very broad uninformative priors
encompassing the entire range of realistic parameter values and
allowing for a wide variety of galaxy models. In particular, the prior
on $\vvir$ is based on the velocity dispersion-halo virial mass
relationship as determined by, e.g., \citet*{Bundy2007} for massive
ellipticals, allowing for a broad standard deviation.

\begin{table}
  \caption{Summary of the adopted priors and of the posterior PDFs
    inferred from the combined lensing and dynamics analysis.}
\begin{small}
    \label{table:par} 
    \begin{tabular}{l@{\quad}l@{\qquad}l@{\qquad}l@{\qquad}l}
      \hline
      parameter              & prior                   & posterior                  & prior                   & posterior                 \\
                             & J0936                   & J0936                      & J0912                   & J0912                     \\
      \hline                                                                                                                                                                   
      $\vvir/\kms$           & $\mathcal{N}$(350, 150) & $ 124^{+160}_{-88} $       & $\mathcal{N}$(475, 200) & $ 470^{+160}_{-130}     $ \\
      \noalign{\smallskip}                                                                                                                                                     
      $\slope$               & U(0, 2)                 & $ 0.92^{+0.72}_{-0.64} $   & U(0, 2)                 & $ 0.46^{+0.41}_{-0.30}  $ \\
      \noalign{\smallskip}                                                                                                                                                     
      $\cg$                  & U(0, 50)                & $ 6.3^{+18.4}_{-4.1} $     & U(0, 50)                & $ 7.2^{+3.8}_{-2.7}     $ \\
      \noalign{\smallskip}                                                                                                                                                     
      $\qh$                  & L$\mathcal{N}$(1, 0.3)  & $ 0.93^{+0.25}_{-0.18} $   & L$\mathcal{N}$(1, 0.3)  & $ 0.54^{+0.10}_{-0.08}  $ \\
      \noalign{\smallskip}                                                                                                                                                     
      $\Ms/10^{11} \Msun$    & U(0, 10)                & $ 3.31^{+0.18}_{-0.30} $   & U(0, 35)                & $ 10.28^{+0.57}_{-0.62} $ \\
      \noalign{\smallskip}                                                                                                                                                     
      $b$                    & U(0, 5)                 & $ 0.89^{+0.37}_{-0.31} $   & U(0, 5)                 & $ 1.94^{+0.24}_{-0.21}  $ \\
      \hline
    \end{tabular}

    In the prior columns, $U(a,b)$ denotes a uniform distribution over
    the open interval $(a, b)$. $\mathcal{N}(a,b)$ and
    $L\mathcal{N}(a,b)$ denote a normal and lognormal distribution,
    respectively, with $a$ being the central value for the variable
    and $b$ the standard deviation (for the log of the variable, in
    the case of $L\mathcal{N}$). In the posterior columns we list, for
    each parameter, the median value of the corresponding marginalized
    posterior PDF and the uncertainty quantified by taking the 68\%
    credible interval (i.e., the 16th and 84th percentiles). See
    Sect.~\ref{sec:mass-model} for a description of the six free model
    parameters.
\end{small}
\end{table}



\section{Results and discussion}
\label{sec:results}

\begin{figure*}
  \centering

  \resizebox{0.84\hsize}{!}{\includegraphics[angle=-90]
    {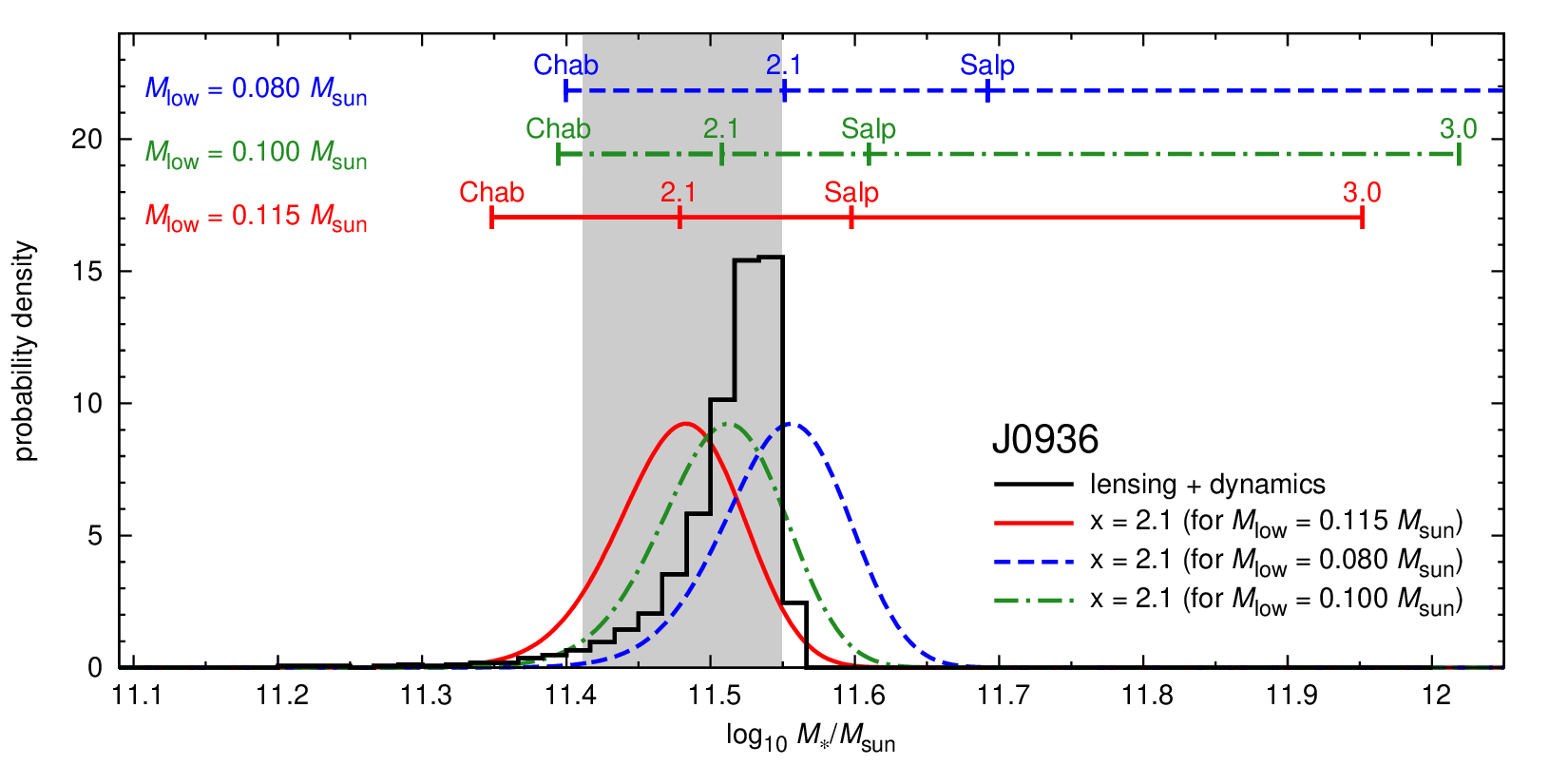}}

  \vspace{-0.1cm}

  \resizebox{0.84\hsize}{!}{\includegraphics[angle=-90]
    {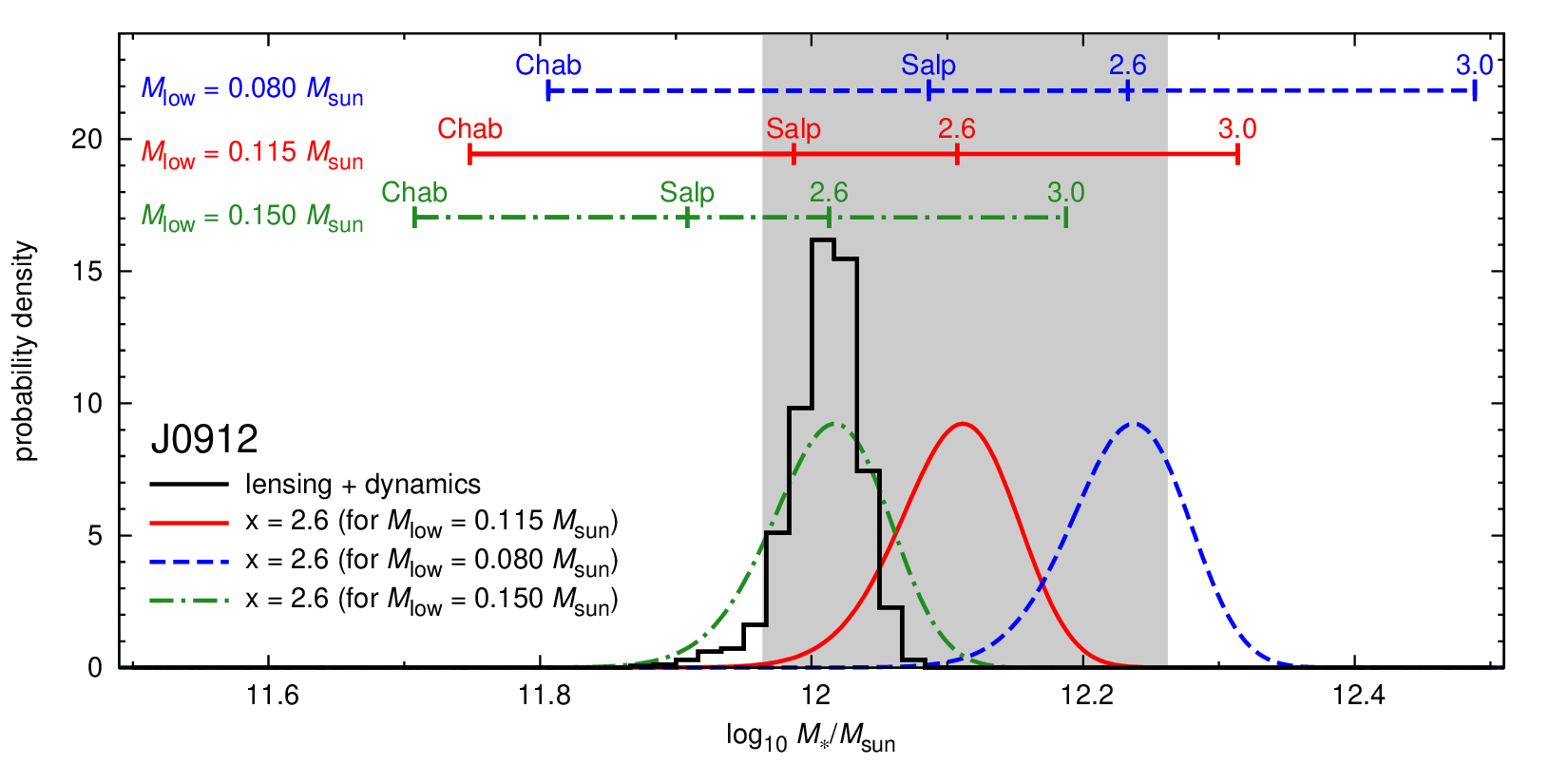}}

  \caption{Comparison of the total stellar masses~$M_{\star}$ inferred
    from the combined lensing and dynamics analysis (histogram) with
    the~$M_{\star}$ determined from the spectroscopic SSP modelling of
    line-strength indices, for different values of the IMF slope
    (indicated above the tics on the horizontal bars). The solid red
    curve and the grey band denote, respectively, $M_{\star}$ for the
    MAP slope and the corresponding 1-sigma uncertainties around that
    value (i.e., $\simfJ{J0936} = 2.10 \pm 0.15$ in the left panel and
    $\simfJ{J0912} = 2.60 \pm 0.30$ in the right panel), derived by
    adopting the low-mass IMF cut-off $\Mcutoff = 0.115 \Msun$ from
    DSEP. The dashed blue curve shows~$M_{\star}$ obtained for the MAP
    slope when we adopt the value $\Mcutoff = 0.080 \Msun$ assumed by
    CvD12. The dash-dotted green curve gives the inference
    for~$M_{\star}$ when using the MAP slope and the best-fit
    $\Mcutoff$ value.}
  \label{fig:IMF}
\end{figure*}

In this Letter we focus on the inferences for the total stellar mass
of the two analyzed systems. The posterior PDFs for the quantity
$\logM$ are shown in Fig.~\ref{fig:IMF} as histograms. The values are
tightly constrained: we find $\logM = 11.52^{+0.02}_{-0.04}$ for J0936
and $\logM = 12.01^{+0.03}_{-0.03}$ for J0912. The corresponding dark
matter fractions within one effective radius~$\Reff$ are $\fDM =
0.06^{+0.10}_{-0.05}$ and $\fDM = 0.18^{+0.08}_{-0.08}$,
respectively. These values are in agreement within the uncertainties
with the findings from the most accurate dynamical studies of local
ellipticals, which measure median $\fDM \approx 0.10 - 0.25$
\citep[see][]{Williams2009, Thomas2011, Cappellari2013-atlasXV}. We
note that J0936 is at the lower limit of these values and is
marginally consistent with having no dark matter contribution within
its inner regions.

From the SSP analysis described in Sect.~\ref{sec:SSP} we obtain a
direct measurement of the IMF slope for the two systems, inferring
$\simfJ{J0936} = 2.10 \pm 0.15$ and $\simfJ{J0912} = 2.60 \pm
0.30$. From these values we derive the total~$M_{\star}$ of the two
objects and we compare them with the corresponding~$M_{\star}$
independently determined from the lensing and dynamics study, as
illustrated in Figure~\ref{fig:IMF}. As clearly seen, the stellar
masses inferred from the two methods are consistent within the
one-sigma uncertainties for both galaxies.

The first result is that both galaxies are consistent with a
Salpeter-like IMF: for J0912, $\simf = 2.35$ is within the 1-sigma
uncertainties derived from the SSP analysis, and is actually preferred
over the \emph{maximum a posteriori} (MAP) slope $\simfJ{J0912} = 2.6$
with an~$84\%$ probability (Bayes factor $B = 5.2$). For J0936, a
Salpeter IMF is slightly disfavoured with respect to the MAP slope
$\simfJ{J0936} = 2.1$, albeit only marginally ($B = 0.40$). In
general, even though we see a weak steepening of the best-fit IMF when
going from the slightly sub-Salpeter J0936 ($\simf \simeq 2.2$) to
J0912 (best-fit slope~$\simf \simeq 2.4$), there is no compelling
indication of an IMF slope trend with galaxy mass from this data~set.

A Chabrier IMF, on the other hand, is ruled out for both systems with
a high degree of confidence.  A Salpeter-like IMF is preferred over a
Chabrier-like one at $89\%$ probability ($B = 8$) for J0936, and at
$99\%$ probability ($B = 96$) for J0912, corresponding in the second
case to very strong evidence.\footnote{Very similar values of the
  Bayes factors are obtained if we compare the Chabrier IMF with the
  MAP slopes rather than with Salpeter.} This confirms the findings of
several studies of ETGs, using a variety of methods
\citep[e.g.][]{Grillo2009, Spiniello2011, Cappellari2012}.

We note that the SSP results are obtained under the assumption of a
single power-law IMF.  A broken power-law (as proposed by
\citealt{Kroupa2001}) or a log-normal IMF (as in
\citealt{Chabrier2003}), both described by two different slopes for
different mass ranges (``tapered IMFs''), will be considered in future
works.  In particular, we plan to apply the method presented here for
the analysis of the entire XLENS sample (Spiniello et al., in prep.)
testing different assumptions for the IMF with the value of $\Mcutoff$
as well as the peak mass characterizing a tapered IMF as free
parameters.

Some works based on absorption-line strengths (e.g.,
\citealt{Ferreras2013}) have suggested that the most massive ETGs
($\sigma \gtrsim 200 \kms$) might have increasingly bottom-heavy IMFs,
with slopes significantly steeper than Salpeter, i.e.~$\simf = 3.0 -
3.5$.
Even though one might have concluded from the inferences drawn from
the spectroscopic SSP analysis alone that the massive galaxy J0912 is
consistent with an $\simf = 3.0$ super-Salpeter IMF, that scenario is
clearly ruled out when also the results of the combined L\&D analysis
are brought into the picture.
In fact, for J0912 (and even more so for J0936) a Salpeter IMF is
favoured over a steep $\simf = 3.0$ IMF by a Bayes factor~$B > 1000$,
corresponding to decisive evidence.\footnote{It has been argued that
  $B > 1000$ is high enough to be used as conclusive forensic evidence
  in criminal trials, cf.~\citet{Kass-Raftery1995} and references
  therein.} More intuitively, adopting a super-Salpeter IMF results in
a stellar mass exceeding the galaxy total mass predicted by the
combined model, and therefore this scenario must be ruled out as
unphysical.

Finally, the approach presented in this Letter allows for the first
time a resolution of the long-standing problem of constraining the
low-mass cut-off for the IMF. In previous studies, $\Mcutoff$ has been
treated as a fully unconstrained parameter, despite being critical to
determine~$\MoLs$. Stars with masses below $\sim 0.15 \, \Msun$ have
very little effect on the spectral lines in the optical and
near-infrared for any assumed IMF slope (CvD12) but they give a
non-negligible contribution to the total mass budget of the system
\citep{Worthey1994}.  Therefore, it is impossible to determine
$\Mcutoff$ from spectroscopic studies alone, since varying this
parameter does not have any real impact on the spectra and on the
line-index measurements. Luckily, this degeneracy can be broken when
adding the information from the combined L\&D analysis.
The analysis conducted so far in this Letter has assumed the cut-off
mass value $\Mcutoff^{\mathrm{D}} = 0.115 \, \Msun$ assumed by
DSEP. However, when adopting a lower value $\Mcutoff^{\mathrm{H}} =
0.08 \, \Msun$, corresponding to the hydrogen burning limit (as
suggested by, e.g., CvD12), the inferred total stellar mass increases
significantly for the steeper IMF slopes, i.e.\ around~$25\%$ for
$\simf = 2.35$ (Salpeter) and~$50\%$ for $\simf = 3.0$. For J0936 the
constraints on~$M_{\star}$ are not good enough to discriminate between
the two possibilities; on the other hand, in the case of J0912
(considering the MAP slope $\simfJ{J0912} = 2.6$), there is decisive
evidence in favour of $\Mcutoff^{\mathrm{D}}$ with $B = 401$,
corresponding to~$99.8\%$ probability.

\begin{figure}

\centering
\resizebox{0.97\hsize}{!}{\includegraphics{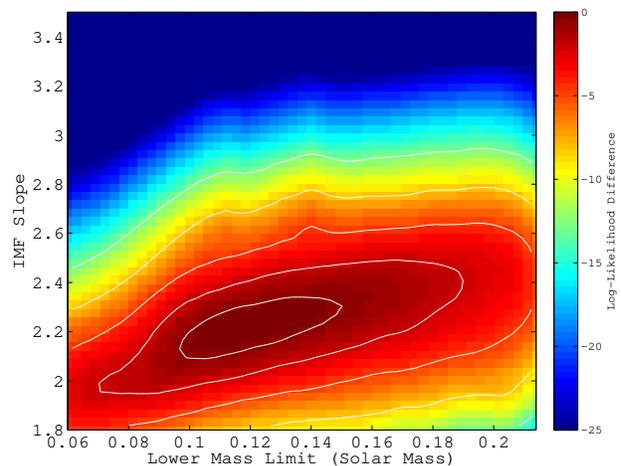}}
\caption{Joint two-dimensional posterior PDF of the IMF slope
  and~$\Mcutoff$ for the two galaxies using both L\&D and SSP
  constraints. The contours denote the equivalent of $\exp(-\chi^2/2)$
  for $\chi = 1, 2, 3, 4$ and $5$.}
 \label{fig:joint_final}

\end{figure}

We combine the results of the L\&D and SSP analyses of both galaxies
to derive the joint inferences on the IMF slope and cut-off.  To this
purpose, we use the DSEP isochrones extrapolated down to $0.06 \Msun$
to compute the V-band $\MoLs$ for different values of the IMF slope
and $\Mcutoff$ for both ETGs; these $\MoLs$ are used to convert the
observed luminosities into stellar masses. The probability density for
this stellar mass is then determined from the posterior PDF of the
L\&D analysis (see Fig.~\ref{fig:IMF}), which acts as prior at this
level of inference. In addition, we assume a Gaussian PDF for the IMF
slope, as determined from the SSP modelling in this paper, and
similarly a Gaussian PDF for the ETG luminosity, determined by Auger
et al (2009), which both act as priors. We adopt a flat prior U(0.06,
0.25) on $\Mcutoff/\Msun$. We then run a Markov Chain Monte Carlo with
$10^5$ samples drawn by varying~$\simf$, $\Mcutoff$ and ETG
luminosity, to sample the joint lensing, dynamics and SSP
posterior. Finally we combine the posteriors derived from the two
galaxies, and marginalize over luminosity to obtain the joint 2-D
posterior PDF, which is shown in Fig.\ref{fig:joint_final}.  The
resulting marginalized MAP values are $\simf = 2.22 \pm 0.14$ for the
IMF slope, and a low-mass cut-off $\Mcutoff = 0.13 \pm 0.03 \, \Msun$.


\section{Conclusions}
\label{sec:conclude}

In this work, for the first time, we compare the results of a
state-of-the-art combined lensing and dynamics analysis of two
early-type galaxies with the corresponding inferences from a
spectroscopic SSP study of line-strength indices. This is made
possible by the exquisite quality of the data sets available for the
two systems, J0936 and J0912, including \emph{HST} high-resolution
imaging and high signal-to-noise X-Shooter spectra.

We model simultaneously the lensing and kinematic data sets by
adopting a flexible, axisymmetric, two-component mass distribution for
the dark halo and the luminous profile (both self-gravitating), and
solving the Jeans anisotropic equations. We derive tight constraints
on the total stellar mass and dark matter fractions (within one
$\Reff$) of the two galaxies, i.e. $\logM = 11.52^{+0.02}_{-0.04}$
(and $\fDM = 0.06^{+0.10}_{-0.05}$) for J0936 and $\logM =
12.01^{+0.03}_{-0.03}$ (and $\fDM = 0.18^{+0.08}_{-0.08}$) for J0912,
independent of the IMF. These dark matter contents are in agreement,
within the uncertainties, with the findings of dynamical studies of
local ellipticals \citep[e.g.][]{Cappellari2013-atlasXV}. By comparing
the posterior PDFs for $M_{\star}$ derived from this combined analysis
with the completely independent results from the spectroscopic SSP
study of the two systems, under the assumption of a single power-law
IMF, we draw the following conclusions on the properties of their
IMFs:
\begin{enumerate}
\item The modelling of the line-strength indices (adopting solar
  metallicity and a cut-off mass $\Mcutoff = 0.115\,\Msun$) provides a
  direct determination of the IMF slopes for the two lenses, namely
  $\simfJ{J0936} = 2.10 \pm 0.15$ and $\simfJ{J0912} = 2.60 \pm 0.30$.
  The total stellar masses inferred from this method are completely
  consistent within the one-sigma uncertainties with the results
  obtained from the combined lensing and dynamics analysis, which is a
  fully independent approach that makes no assumptions on the IMF.
\item For both galaxies, an IMF with slope close to Salpeter ($\simf =
  2.35$) provides a much better description than a Chabrier profile:
  statistical analysis based on the calculation of evidence ratios
  (i.e., Bayes factor) shows that a Salpeter IMF model is preferred
  over a Chabrier one with~$89\%$ and~$99\%$ probability for J0936 and
  J0912, respectively.
\item Bottom-heavy IMFs significantly steeper than Salpeter, i.e. with
  slopes $\simf \ge 3.0$, are ruled out with decisive evidence (Bayes
  factor $B > 1000$) for both galaxies. The values for~$M_{\star}$
  implied by these super-Salpeter IMFs unphysically exceed the total
  galaxy mass derived from the combined lensing and dynamics
  analysis. This result contradicts the findings from absorption line
  spectroscopy studies (see e.g. \citealt{Ferreras2013}) which support
  IMF slopes $\simf = 3.0 - 3.5$ for massive ETGs such as those
  considered here ($\sigma_{\mathrm{J0936}} = 243 \pm 11$ $\kms$ and
  $\sigma_{\mathrm{J0912}} = 326 \pm 12$ $\kms$).
\item For the system J0912, a low-mass IMF cut-off of $\Mcutoff = 0.08
  \, \Msun$ (corresponding to the hydrogen burning limit) is ruled out
  with a probability of~$99.8\%$ in favor of the value
  $\Mcutoff^{\mathrm{D}} = 0.115\,\Msun$ assumed by DSEP. For J0936,
  instead, it is not possible to reliably discriminate between the two
  values.
\item By combining the L\&D and SSP inferences of the two lenses, we
  obtain the marginalized MAP values $\simf = 2.22 \pm 0.14$ for the
  IMF slope and $\Mcutoff = 0.13 \pm 0.03 \, \Msun$, consistent with a
  Salpeter-like IMF and a low-mass cut-off slightly higher than the
  hydrogen burning limit.

\end{enumerate}

\section*{Acknowledgments}

%
%
The Dark Cosmology Centre is funded by the DNRF.
CS acknowledges support from an Ubbo Emmius Fellowship.
TT acknowledges support from a Packard Research Fellowship.



\label{lastpage}

\clearpage

\end{document}